\pdfoutput=1
\documentclass[11pt,a4paper]{article}
\usepackage[T1]{fontenc}
\usepackage{lmodern}
\usepackage[margin=27mm]{geometry}
\usepackage{amsmath,amssymb}
\usepackage{graphicx}
\usepackage[numbers,square,sort&compress]{natbib}
\usepackage[hidelinks]{hyperref}
\usepackage{doi}
\title{Refinements to precision-information imputation in meta-analysis of survival rates: Boundary correction and cross-study borrowing}
\author{\parbox{0.96\linewidth}{\centering
  Kazushi Maruo\textsuperscript{1}, Yusuke Yamaguchi\textsuperscript{2},
  Ryota Ishii\textsuperscript{1},\\
  Hisashi Noma\textsuperscript{3}, and Masahiko Gosho\textsuperscript{1}\\[0.8ex]
  {\small
  \textsuperscript{1} Department of Biostatistics, Institute of Medicine, University of Tsukuba, Ibaraki, Japan\\
  \textsuperscript{2} Biostatistics, Data Science, Astellas Pharma Global Development, Inc., Northbrook, IL, USA\\
  \textsuperscript{3} Department of Data Science, The Institute of Statistical Mathematics, Tokyo, Japan\\
  Corresponding author Kazushi Maruo (\href{mailto:kazushi.maruo@gmail.com}{\nolinkurl{kazushi.maruo@gmail.com}})
  }
}}
\date{}
\hypersetup{
  pdftitle={Refinements to precision-information imputation in meta-analysis of survival rates: Boundary correction and cross-study borrowing},
  pdfauthor={Kazushi Maruo, Yusuke Yamaguchi, Ryota Ishii, Hisashi Noma, Masahiko Gosho}
}

\begin{document}
\maketitle

\begin{abstract}
Meta-analyses of survival rates require precision information, such as standard errors (SEs), which is often unreported. We previously proposed imputing missing SEs from reported survival rates, sample sizes, and risk-set sizes. The original method has two limitations. First, it replaces Kaplan--Meier estimates of 1 with 0.99 regardless of sample size. Second, when the risk-set size is unavailable, it uses the total event count as a fallback. The count is often unreported, and the fallback can also substantially overestimate SEs. We propose a sample-size-dependent boundary correction based on a pseudo-event and a method that borrows time-specific event information from studies with available risk-set sizes. In simulations, the boundary correction was more stable than fixed replacement, and borrowing reduced study-level SE imputation bias. Borrowing generally improved precision by retaining studies with missing SEs that lacked both risk-set sizes and event counts.
\end{abstract}


\section*{Highlights}

\noindent\textbf{What is already known?} Our previous method imputes missing study-level standard errors (SEs) in meta-analyses of survival probabilities. It replaces Kaplan--Meier estimates of 1 with 0.99 regardless of sample size. When risk-set sizes are missing, it relies on total event counts, which are often unreported.

\medskip
\noindent\textbf{What is new?} We replace the fixed 0.99 rule at the upper boundary with a fractional pseudo-event correction and, when the risk-set size is missing, borrow time-specific event information from other studies.

\medskip
\noindent\textbf{Potential impact.} Both refinements retain the simplicity of the original method, require no reconstruction of individual participant data, and extend its stability and applicability.

\section{Introduction}

Meta-analysis of survival probabilities at specified time points requires the survival estimate and its precision, but the latter is not always reported. We previously proposed a simple method that imputes a missing SE from published aggregate information, including the survival probability, sample size, and risk-set size at the evaluation time~\citep{maruo2025}.

Two practical limitations remained. First, a Kaplan--Meier estimate of 1 yields a zero or undefined SE, depending on the analysis scale. The original method therefore replaced 1 with 0.99 regardless of sample size. Second, the original method uses overall event proportions as proxies
for time-specific proportions when risk-set sizes are missing. This can substantially overestimate SEs, and total event counts may be unreported
for composite outcomes such as progression-free survival. We previously mentioned using the mean of other studies' overall event proportions when a study's own was unavailable~\citep{maruo2025}. This unevaluated fallback can still yield a value that differs from the time-specific proportion.

We therefore propose and evaluate two refinements. A fractional pseudo-event yields a study-size-dependent correction for Kaplan--Meier estimates of 1. To impute missing risk-set sizes, we use a weighted mean of time-specific event proportions from studies reporting risk-set sizes at the same evaluation time.
\section{Methods}

\subsection{Original method}

Let $\widehat S_r(t)$ denote the Kaplan--Meier estimate at time $t$ in study $r$ $(r=1,\ldots,R)$. The variance estimators for $\widehat{S}_r(t)$ on the untransformed, log, and log--log scales are
\begin{align*}
\widehat{\mathrm{Var}}\{\widehat S_r(t)\}
  &= \eta_r(t)\widehat S_r(t)^2, \\
\widehat{\mathrm{Var}}\{\log \widehat S_r(t)\}
  &= \eta_r(t), \\
\widehat{\mathrm{Var}}\!\left[\log\{-\log \widehat S_r(t)\}\right]
  &= \frac{\eta_r(t)}{\{\log \widehat S_r(t)\}^2},
\end{align*}
where
\begin{equation*}
\eta_r(t)=\sum_{t_{ir}\le t}\frac{d_{ir}}{n_{ir}(n_{ir}-d_{ir})},
\end{equation*}
and $n_{ir}$ and $d_{ir}$ are the risk-set size and number of events at event time $t_{ir}$.  
Based on the approach of Maruo et al.~\citep{maruo2025}, we approximate this quantity by
\begin{equation*}
\eta_r(t)\simeq p_{rt}\sum_{j=n_{rt}+1}^{n_r}\frac{1}{j(j-1)}= p_{rt}\left\{\frac{1}{n_{rt}}-\frac{1}{n_r}\right\},
\end{equation*}
where $n_r$ is the initial sample size, $n_{rt}$ is the risk-set size at $t$, and $p_{rt}$ is the proportion of events among individuals removed from the risk set by $t$. Its natural estimator is $\widehat p_{rt}=D_{rt}/(n_r-n_{rt})$, where $D_{rt}$ is the cumulative number of events by $t$. Under the assumptions described previously~\citep{maruo2025},
\begin{equation}
\widehat p_{rt}\simeq
\frac{(n_r+n_{rt})\{1-\widehat S_r(t)\}}
     {(n_r-n_{rt})\{1+\widehat S_r(t)\}}.
\label{eq:prt}
\end{equation}
If $n_{rt}$ is unavailable, the original method solves Equation~\eqref{eq:prt} for $n_{rt}$ after replacing the time-specific $p_{rt}$ by the overall event proportion $p_r$.

\subsection{Boundary correction}

When $\widehat S_r(t)=1$ and $n_{rt}$ is observed, we introduce a
fractional pseudo-event $c>0$ by time $t$ and define
\begin{equation*}
\widetilde n_{rt}=n_{rt}-c,
\qquad
\widetilde S_r(t)=1-\frac{2c}{n_r+n_{rt}}.
\end{equation*}
If $n_{rt}=n_r$, then $\widetilde S_r(t)=1-c/n_r$; thus, unlike replacement by 0.99, the correction decreases as study size increases. The corrected $\widetilde S_r(t)$ and $\widetilde n_{rt}$ are then substituted into the Greenwood approximation above.

\subsection{Borrowing time-specific information across studies}

When $n_{rt}$ is unavailable, we borrow time-specific event information from studies for which $n_{rt}$ is observed. Let $A_t$ denote this donor set. We use
\begin{equation*}
\overline p_t=
\frac{\sum_{r\in A_t}(n_r-n_{rt})\widehat p_{rt}}
     {\sum_{r\in A_t}(n_r-n_{rt})}.
\end{equation*}
Under a working binomial model with a common event probability $p_t$ among individuals who have left the risk set by $t$, the conditional variance of $D_{rt}/(n_r-n_{rt})$ given $n_r-n_{rt}$ is $p_t(1-p_t)/(n_r-n_{rt})$.
The inverse-variance weights for these count-based proportions are thus proportional to $n_r-n_{rt}$, motivating the use of the same weights for the approximations in Equation~\eqref{eq:prt}.
For missing $n_{rt}$, let $q$ denote the event proportion used for imputation ($p_r$ or $\overline p_t$). For $0<\widehat S_r(t)<1$, we substitute $q$ for $p_{rt}$ and invert Equation~\eqref{eq:prt}. When
$\widehat S_r(t)=1$ and $q>0$, the initial corrections are $\widetilde n_{rt}=n_r-c/q$ and $\widetilde S_r(t)=1-2c/(n_r+\widetilde n_{rt}+c)$. 
We first raise the risk-set size to at least 2 and then cap it at $n_r-1$ ($n_r-c$ when $\widehat S_r(t)=1$). We re-estimate $p_{rt}$ using this size and the survival estimate, cap $p_{rt}$ at 1, and compute the SE.

\begin{figure}[b!]
\centering
\includegraphics[width=0.95\linewidth]{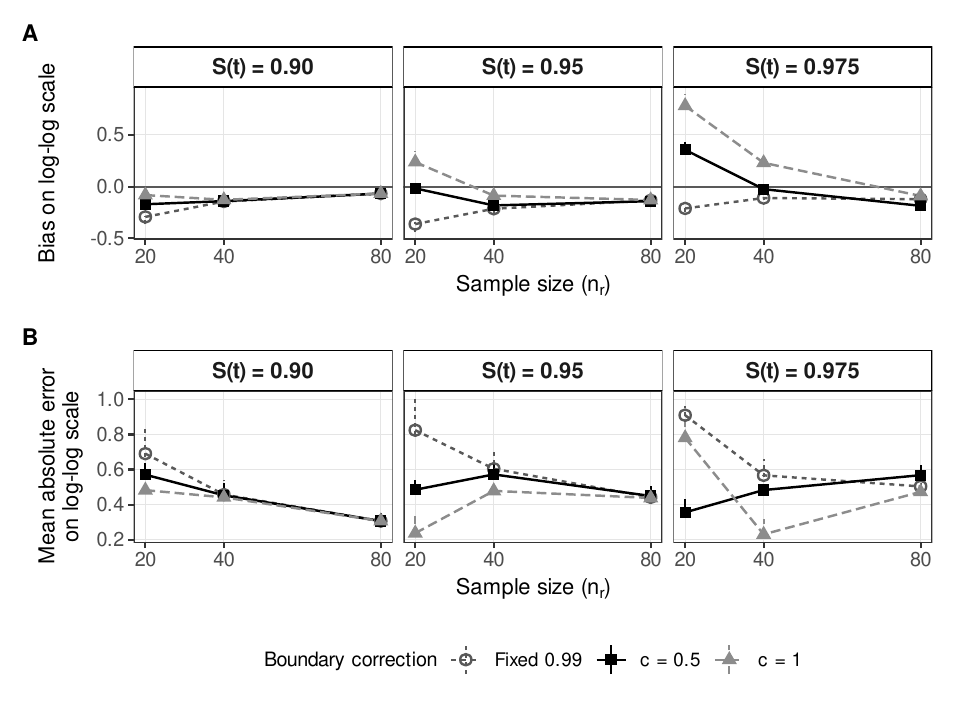}
\caption{Performance of boundary corrections in the study-level simulation. Panel A shows bias and Panel B mean absolute error on the log--log scale. Points/lines summarize medians and vertical ranges summarize the minimum to maximum across the remaining simulation factors}
\label{fig:boundary}
\end{figure}

\begin{figure}[b!]
\centering
\includegraphics[width=0.95\linewidth]{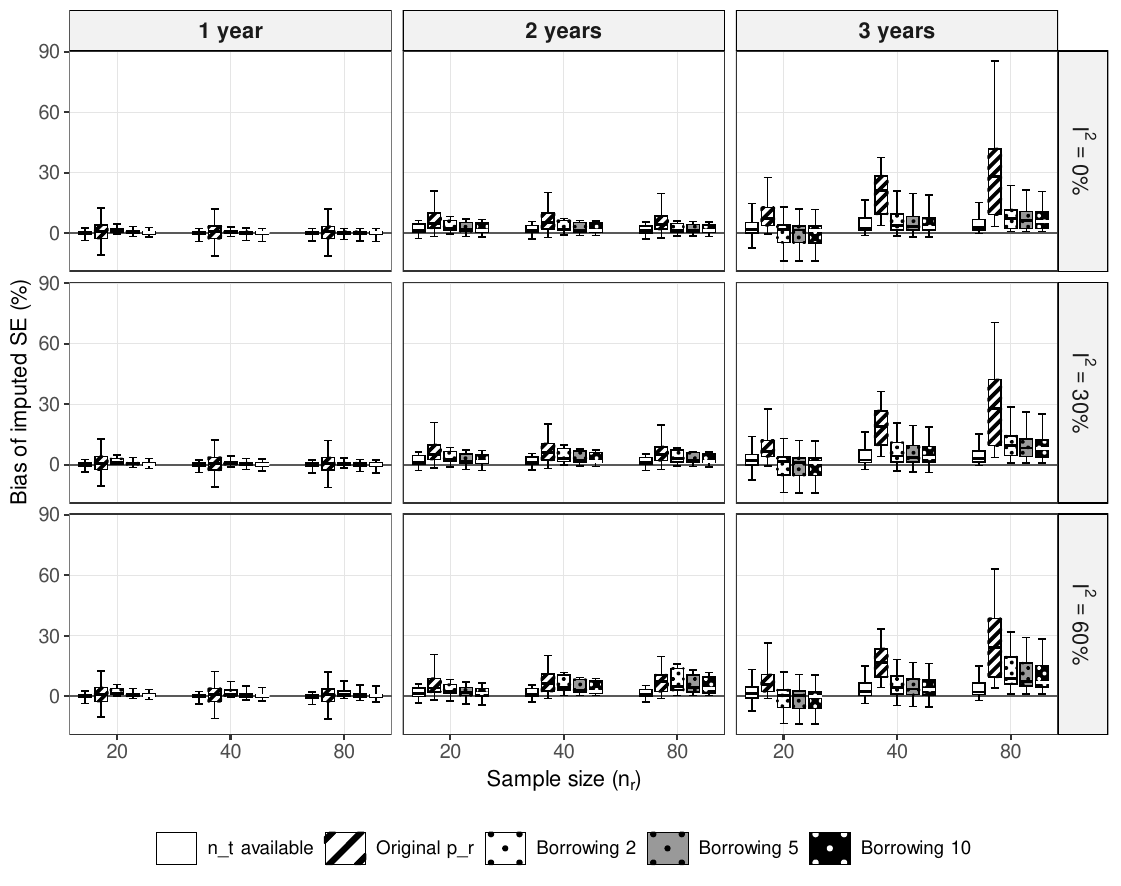}
\caption{Relative bias of imputed standard errors in the study-level borrowing simulation. The ideal condition with observed $n_{rt}$, the original method using $p_r$, and borrowing from 2, 5, or 10 donor studies are compared}
\label{fig:studyborrow}
\end{figure}

\begin{figure}[b!]
\centering
\includegraphics[width=0.95\linewidth]{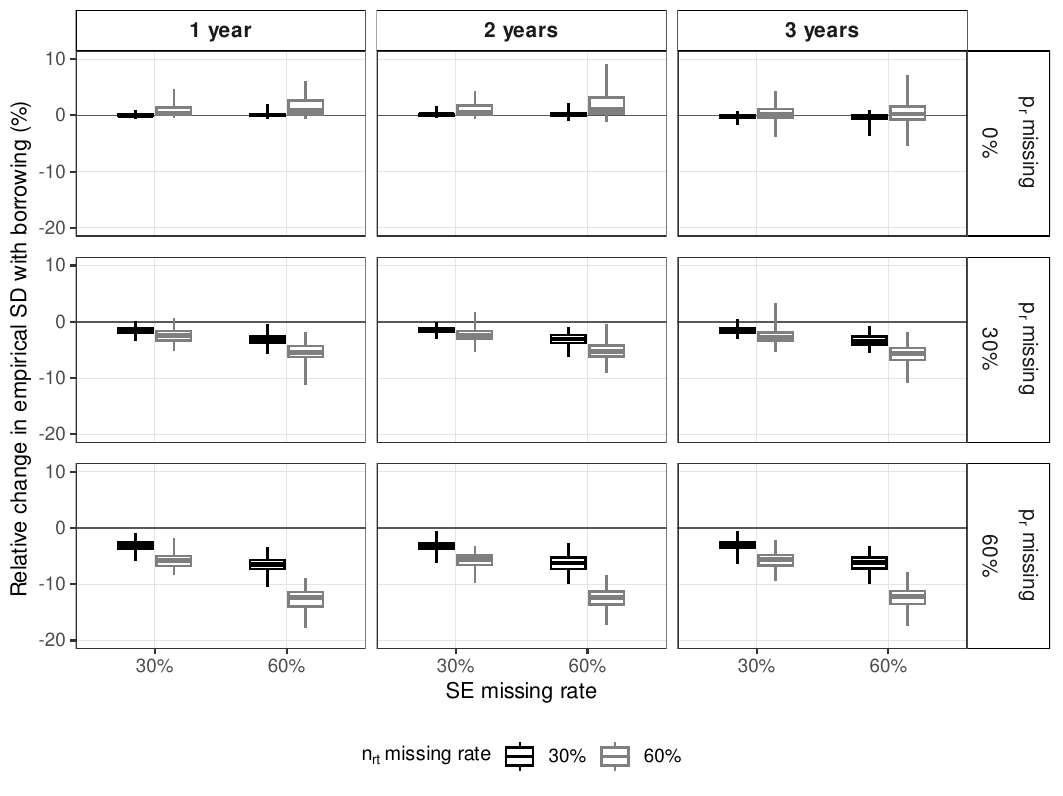}
\caption{Relative change in empirical SD when borrowing replaces the original $p_r$ method in the meta-analysis-level simulation. Negative values indicate improved precision}
\label{fig:meta}
\end{figure}

\section{Simulation evaluation}
We adapted the simulation settings described in our previous study~\citep{maruo2025}.
We evaluated boundary correction in study-level simulations. Event and censoring times followed Weibull distributions with shape parameters 0.5, 1, or 2; censoring proportions were 20\% or 40\%; and sample sizes were 20, 40, or 80. The event-time scale was calibrated so that 2-year survival was 0.5. Across 54 conditions and 10,000 replicates per condition, we estimated survival at time points corresponding to true survival probabilities of 0.90, 0.95, and 0.975. Fixed 0.99 replacement was compared with $c=0.5$ and $c=1$ using bias and mean absolute error (MAE) on the log--log scale. 
Figure~\ref{fig:boundary} shows the results. In small studies, fixed replacement showed negative bias, and $c=1$
overcorrected when true survival was close to 1. We used $c=0.5$ thereafter to limit this
overcorrection.

We next evaluated borrowing when $n_{rt}$ was missing. The same basic settings were used, with between-study heterogeneity corresponding to $I^2=0\%$, 30\%, or 60\%. Across 162 conditions and 10,000 replicates per condition, 1-, 2-, and 3-year survival probabilities were evaluated. We compared SE imputation using observed \(n_{rt}\) (benchmark), the original \(p_r\) method, and the proposed borrowing method (2, 5, or 10 donors). Figure~\ref{fig:studyborrow} shows the results. Relative SE bias was generally smaller with borrowing than with the original method, especially at 3 years. Additional improvement from increasing the number of donors was modest.

In the meta-analysis-level simulation, the number of studies was 5, 10, or 20, and each study size was sampled from $\mathrm{TLN}(\log 20,1;10,100)$, a log-normal distribution truncated to 10--100. Missingness was generated at rates of 30\% or 60\% for SE and $n_{rt}$ and 0\%, 30\%, or 60\% for $p_r$. The original $p_r$ method excluded studies missing all three quantities. Across 1,944 conditions and 1,000 replicates per condition, 1-, 2-, and 3-year survival probabilities were pooled using random-effects meta-analysis with REML estimation and Hartung--Knapp confidence intervals~\citep{hartung1999,knapp2003} in the R package \texttt{meta}~\citep{balduzzi2019}. 
We assessed precision gains from borrowing using the percentage change in
empirical SD of pooled estimates relative to the original $p_r$
method on the log-log transformed scale. We also evaluated bias and 95\% interval coverage.
Borrowing generally improved precision at $p_r$ missingness rates of 30\% and 60\%, with little change when $p_r$ was available for all studies (Figure~\ref{fig:meta}). Median coverage of the 95\% confidence intervals was 94.5\% for both
methods, with ranges of 90.0--97.1\% for the $p_r$ method and 88.8--97.0\% for borrowing. 
Both methods showed little bias (see the supplementary results).

\section{Discussion}

In this study, we proposed and evaluated two refinements to SE imputation in meta-analyses of survival probabilities.

A Kaplan--Meier estimate of 1 provides stronger evidence of high survival when more participants remain at risk. The fractional pseudo-event correction makes smaller adjustments as sample size or risk-set size increases. We chose a pseudo-event count of 0.5 to limit the overcorrection seen with one pseudo-event in small studies when true survival was close to 1.

Borrowing time-specific event proportions generally reduced study-level SE imputation bias. Gains in pooled precision were most apparent when some studies lacked overall event proportions. With suitable donors, borrowing can retain studies missing SEs, risk-set sizes, and overall event proportions. When overall proportions were available for all studies, borrowing generally had little effect on pooled precision and sometimes reduced it under substantial between-study heterogeneity.

Both refinements use reported aggregate data, avoiding individual participant data reconstruction and Kaplan--Meier curve digitization. They preserve the original method's simplicity while improving stability and extending SE imputation to more studies. Both refinements are implemented in the R package \texttt{metaSurvMissCI} (version 0.2.0).

\paragraph{Author contributions}
K.M. led conceptualization, methodology, formal analysis, software, and writing of the original draft. Y.Y. and R.I. contributed to software in supporting roles. H.N. and M.G. contributed equally to supervision. M.G. contributed to funding acquisition. All authors contributed equally to reviewing and editing the manuscript.

\paragraph{Competing interest statement}
The authors declare that no competing interests exist.

\paragraph{Data availability statement}
No empirical data were used. Simulation results and code are available as ancillary files accompanying this preprint.

\paragraph{Ethics statement}
Ethics approval and patient consent were not applicable to this simulation study.

\paragraph{Funding statement}
This study was supported by JSPS KAKENHI Grant Numbers JP24K02662 and JP26K02873.

\paragraph{Supplementary Material}
The ancillary files include \texttt{RSM\_simulation\_results.xlsx} (simulation results)
and \texttt{Simulation\_public.zip} (simulation and plotting code).

\paragraph{Acknowledgements}
In November 2026, we used ChatGPT (OpenAI, \url{https://chatgpt.com})
to draft and revise parts of the manuscript, develop R scripts for simulations
and figures, and check simulation summaries.

\bibliographystyle{unsrtnat}
\bibliography{survse_imp2_arxiv_references}

@article{maruo2025,
  author  = {Maruo, Kazushi and Yamaguchi, Yusuke and Ishii, Ryota and Noma, Hisashi and Gosho, Masahiko},
  title   = {Simple imputation method for meta-analysis of survival rates when precision information is missing},
  journal = {Research Synthesis Methods},
  year    = {2025},
  volume  = {16},
  number  = {6},
  pages   = {937--952},
  doi     = {10.1017/rsm.2025.10024}
}

@article{hartung1999,
  author  = {Hartung, Joachim},
  title   = {An Alternative Method for Meta-Analysis},
  journal = {Biometrical Journal},
  year    = {1999},
  volume  = {41},
  number  = {8},
  pages   = {901--916},
  doi     = {10.1002/(SICI)1521-4036(199912)41:8<901::AID-BIMJ901>3.0.CO;2-W}
}

@article{knapp2003,
  author  = {Knapp, Guido and Hartung, Joachim},
  title   = {Improved tests for a random effects meta-regression with a single covariate},
  journal = {Statistics in Medicine},
  year    = {2003},
  volume  = {22},
  number  = {17},
  pages   = {2693--2710},
  doi     = {10.1002/sim.1482}
}

@article{balduzzi2019,
  author  = {Balduzzi, Sara and R{\"u}cker, Gerta and Schwarzer, Guido},
  title   = {How to perform a meta-analysis with {R}: a practical tutorial},
  journal = {Evidence-Based Mental Health},
  year    = {2019},
  volume  = {22},
  number  = {4},
  pages   = {153--160},
  doi     = {10.1136/ebmental-2019-300117}
}

\end{document}